\documentclass[useAMS,usenatbib,fleqn]{mn2e}  
\usepackage{mathptmx}
\usepackage{graphicx}
\usepackage{amsmath}
\usepackage{amssymb}
\usepackage{psfig}
\usepackage{multirow}
\usepackage{cases}
\usepackage{subfigure}
\usepackage{float}
\usepackage[caption = false]{subfig}
\usepackage[abs]{overpic}

\def\Msun{\hbox{$\thinspace M_{\odot}$}}

\def\xmm{{\it XMM-Newton}}
\newcommand{\xmmn}{{\it XMM-Newton~\/}}

\def\Mbh{\hbox{$M_{\rmn{BH}}$}}

\def\Ledd{\hbox{$L_{\rm Edd}$}}

\def\xmm{{\it XMM-Newton}}
\newcommand{\rej}{RE J1034+396}

\def\gsim{\mathrel{\hbox{\rlap{\hbox{\lower4pt\hbox{$\sim$}}}\hbox{$>$}}}}
\def\lsim{\mathrel{\hbox{\rlap{\hbox{\lower4pt\hbox{$\sim$}}}\hbox{$<$}}}}

   \title[The QPO in \rej]{Detection of a QPO in 5 \xmmn observations of \rej}
   \author[W. N. Alston et al.]{W. N. Alston\thanks{wna@ast.cam.ac.uk},
        J. Markevi\v{c}i\={u}t\.{e}, E. Kara, A. C. Fabian and M. Middleton \\
        Institute of Astronomy, Madingley Rd, Cambridge, CB3 0HA.
}

\date{Accepted 2014 July 28. Received 2014 July 28; in original form 2014 June 24}
 
\pagerange{\pageref{firstpage}--\pageref{lastpage}} \pubyear{2014}

\begin{document}
\label{firstpage}
\maketitle

\begin{abstract}
The X-ray quasi-periodic oscillation (QPO) at $\sim 2.6 \times 10^{-4}$\,Hz in \rej~has been robustly observed in the 0.2--10 keV band of one $\sim 90$\,ks \xmmn observation, but was not detected in subsequent observations.  Here, we investigate the power spectral density (PSD) of 8 archival \xmmn observations of \rej, and search for the presence of QPOs in three energy bands: soft (0.3--0.8\,keV); hard (1.0--4.0\,keV) and total (0.2--10.0\,keV).  We find a significant detection of a QPO feature in the hard-band PSD of five low flux/spectrally-harder observations.  The QPO frequency has remained persistent at $\sim 2.6 \times 10^{-4}$\,Hz over $\sim 4$ years of observations, though it is no longer detected in the soft band, except in the original observation.  This result increases the duration where the QPO is present by a factor of three (now $\sim 250$\,ks), allowing for a better understanding of the QPO phenomenon observed in both Active Galactic Nuclei (AGN) and black hole X-ray Binaries (BHBs).
\end{abstract}


%

\section{Introduction}
\label{sect:intro}

Black hole accretion powers Active Galactic Nuclei (AGN; $\Mbh \gsim 10^{6} \Msun$) and Black Hole X-ray binaries (BHBs; $\Mbh \sim 10 \Msun$).  If the inner accretion flow is dominated by strong gravity then many important aspects of the underlying accretion physics are expected to be scale invariant (e.g. \citealt{shaksuny73}; \citealt{mushotzky93}).  The black hole accretion process should be entirely described by the mass and spin, but their observational appearance will also be determined by the mass accretion rate, geometry, and other system parameters.

Several BHBs display high-frequency quasi-periodic oscillations (HFQPOs) at $\nu \gsim 40$\,Hz (e.g. \citealt{Remillard1999}; \citealt{Strohmayer2001}; \citealt{remillard2002, remillard2003prec}; \citealt{vanderklis2006}, see \citealt{RemillardMcClintock06} for a review).  These are the fastest coherent features observed in BHBs, with frequencies close to the Keplerian frequency of the innermost stable circular orbit (ISCO).  With an origin in the strongly-curved spacetime close to the black hole, HFQPOs should therefore carry information about the two fundamental parameters of black holes: mass and spin.  A scale invariance of the accretion process implies that HFQPOs should also be present in AGN.

The first robust AGN HFQPO detection came from the Seyfert galaxy \rej, with a $\sim 1$\,hr periodicity \citep[][hereafter G08]{gierlinski08}.  Later, a $\sim 200$\,s QPO was reported during a tidal disruption event in Swift J164449.3+573451 (\citealt{ReisETAL13}).  More recently, a $\sim 3.8$\,hr QPO was confirmed in 2XMM J123103.2+110648 (\citealt{LinETAL13}), which the authors associated with the low-frequency QPO phenomenon due to the low black hole mass ($\Mbh \sim 10^{5} \Msun$) and 50 per cent rms variability.  A systematic study of AGN power spectra was presented in \citet{GonzalezMartinVaughan12}, where no significant QPOs were detected, except the known QPO in \rej.

The QPO at $\sim 2.6 \times 10^{-4}$\,Hz in \rej~was originally detected in the 0.3--10.0\,keV band from a $\sim 90$\,ks \xmmn observation by G08 and \citet{MiddletonETAL09}.  This feature was confirmed by \citet[][hereafter V10]{vaughan10}, although the detection was claimed to be less significant.  The 0.3--10.0\,keV PSD of four further observations of \rej~was studied in \citet[][hereafter M11]{middletonetal11}, where no evidence for a periodic feature was seen.  However, M11 found tentative evidence for the presence of a QPO in two low flux/spectrally-harder observations, through analysis of the covariance spectra (\citealt{WilkinsonUttley09}), suggesting the QPO is a transient feature.  Further attempts at understanding the QPO in \rej~have limited themselves to the one $\sim 90$\,ks observation (\citealt{CzernyETAL2010}; M10; \citealt{Czerny12}; \citealt{HuETAL14}).

HFQPOs are observed in certain BHB `states' (very high/intermediate), associated with high mass accretion rates (e.g. \citealt{remillard2002}; \citealt{RemillardMcClintock06}).  The high mass accretion rate of \rej~($L_{\rm Bol}/\Ledd \sim 1$) led \citet[][hereafter M10]{MiddletonDone10} and M11 to conclude that the \rej~is an analogue of the 67\,Hz QPO in the super-Eddington BHB GRS 1915+105 (and possibly the same feature is being seen in the black hole candidate IGR J17091--3624 \citealt{Altamirano2012}).  

In this letter we show that the QPO feature is present in all of the low flux/spectrally-harder observations, numbering five in total.  The energy dependence of the QPO has changed over the 4 year period between the observations, yet the QPO frequency remains unchanged at $\sim 2.6 \times 10^{-4}$\,Hz. 

The structure of this letter is as follows: in Section~\ref{sec:obs} we describe the observations and data reduction, in Section~\ref{sec:psd} we present the power spectral analysis and QPO identification, and in Section~\ref{sec:disco} we discuss the results in terms of current QPO mechanisms and the analogies with BHB HFQPOs.\\


\section{Observations and data reduction}
\label{sec:obs}

We make use of the 8 publicly available \xmmn observations of \rej.  For the PSD analysis we are most concerned with long observations (i.e. with durations \textgreater 25\,ks), however we do include the short $\sim 15$\,ks observation taken in 2002.  The details of these observations are shown in Table~\ref{obs_tab}.

We focus only on the EPIC-pn (European Photon Imaging Camera; \citealt{struder01}) data.  The Observation Data Files (ODFs) were processed following standard procedures using the \xmmn\ Science Analysis System (SAS v13.5.0), with filtering conditions {\tt PATTERN} = 0--4 and {\tt FLAG} = 0.  Source light curves were extracted in a 20 arcsec circular region and the background was taken from a large rectangular region on the same chip.  We filtered out soft proton flares using a threshold of $2~{\rm ct~s}^{-1}$ in the 10.0--12.0\,keV background light curve.  We linearly interpolate across any short gaps and add Poisson noise.  The interpolation fraction was typically $\lsim 0.5$ per cent.  The background was visually inspected for rises towards the end of the observations, and we clipped the end of the observations if the background rate became approximately equal to the source rate.  The duration of the clipped light curves that are used in the subsequent analysis are given in Table~\ref{obs_tab} column 4.

The 2002 observation (hereafter Obs0) and the first long observation (hereafter Obs1) were taken in full-window (FW) mode, and as such suffers from pile-up (M11).  For these observations only we use a 40 arcsec extraction region and excise the central 7.5 arcsec for consistency with G08 and M11.  The remaining observations (hereafter Obs2--Obs7) were taken in small-window (SW) mode, and do not suffer the effects of pile-up  (\citealt{ballet99}; \citealt{davis01}).

\begin{figure}
\includegraphics[width=0.48\textwidth,angle=0]{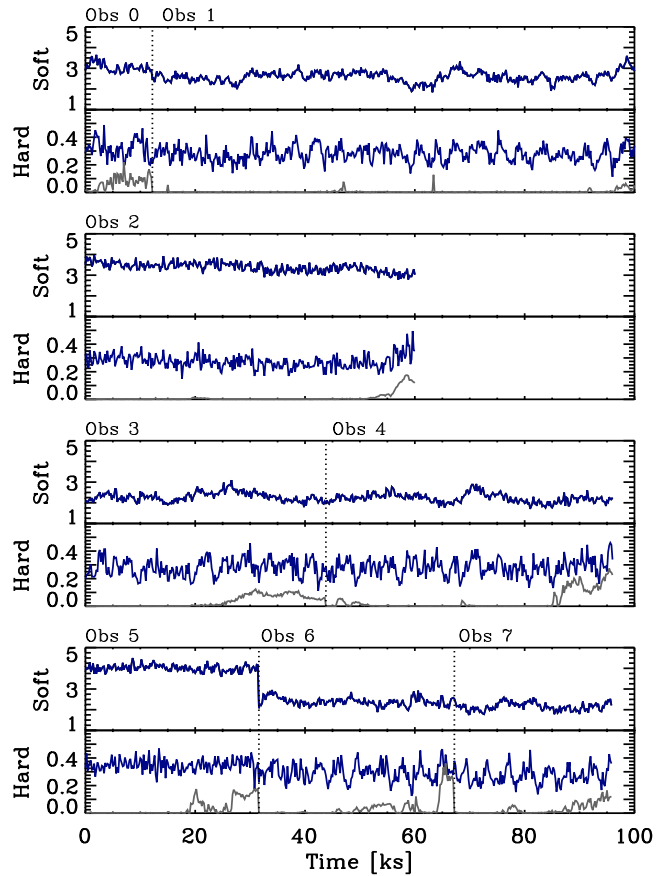}
\caption{Soft (0.3--0.8\,keV) and hard (1.0--4.0\,keV) background-subtracted source (blue) and background (grey) light curves (pre-clipped) for all the \xmmn observations listed in Table~\ref{obs_tab}.}
\label{fig:ltcrv}
\end{figure}

The 0.3--0.8\,keV (soft) and 1.0--4.0\,keV (hard) light curves are shown in Fig.~\ref{fig:ltcrv}, which show distinct spectral changes (M10; M11).  In particular, Obs2 and 5 exhibit a much softer spectra, which can be seen from the hardness ratio, $R = H / (S+H)$, shown in Table~\ref{obs_tab} column 5.  M11 showed that the increase in soft flux in Obs2 was due to an increase in the soft-excess component normalization.  The high frequency variability is significantly reduced in Obs2 (M11).  The mean hard-band count rate is approximately constant over the set of observations.

\begin{table*}
\caption{\textit{XMM-Nexton} observation summary. The columns list (1) Observation number, (2) observation ID, (3) start date of the observation, (4) EPIC-pn observation duration (rounded down to nearest ks), (5) the source count rate in the 0.3--0.8\,keV (S), 1.0--4.0\,keV (H), 0.2--10.0\,keV (T) bands and hardness ratio $R = H / (S+H)$, respectively.}
\begin{tabular}{l c c c c}
\hline
Obs & Obs. ID & Date & Duration  & Count rate\\
No. & & & [Clipped] & S / H / T / R \\
 & & (UT) & (ks) & (ct s$^{-1}$) \\
(1) & (2) &(3) &(4) &(5) \\
\hline
0 & 0109070101 & 2002-05-01 & 15 [11] & 2.81 / 0.30 / 4.85 / 0.10   \\
1 & 0506440101 & 2007-05-31 & 91 [90] & 2.60 / 0.28 / 4.72 / 0.10   \\
2 & 0561580201 & 2009-05-31 & 60 [55] & 3.41 / 0.27 / 5.63 / 0.08   \\
3 & 0655310101 & 2010-05-09 & 44 [41] & 2.31 / 0.28 / 4.12 / 0.11   \\
4 & 0655310201 & 2010-05-11 & 52 [52] & 2.22 / 0.28 / 3.94 / 0.11   \\
5 & 0675440101 & 2011-05-27 & 32 [30] & 4.01 / 0.35 / 6.62 / 0.08   \\
6 & 0675440201 & 2011-05-31 & 36 [35] & 2.39 / 0.29 / 4.18 / 0.11   \\
7 & 0675440301 & 2011-05-07 & 29 [27] & 2.14 / 0.27 / 3.86 / 0.11   \\
\hline
\end{tabular}
\label{obs_tab}
\end{table*}

\section{Energy-dependent power spectrum}
\label{sec:psd}

\begin{figure}
  \centering
\includegraphics[width=0.35\textwidth,angle=90]{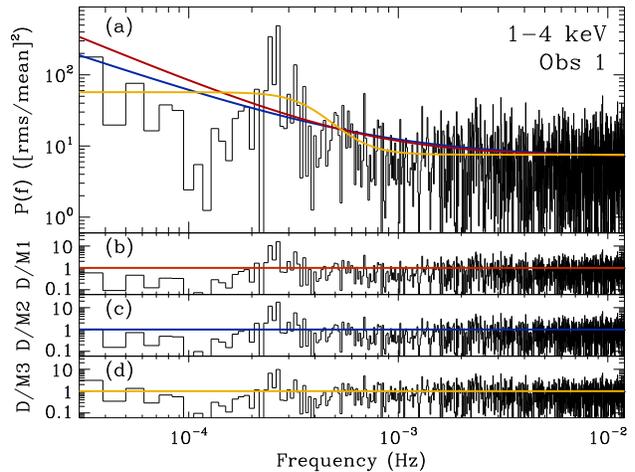}
\caption{Obs1 data and PSD model fits for the 1.0--4.0\,keV (hard) band are shown in panel (a), for Model 1 (red), Model 2 (blue) and Model 3 (yellow).  Panels (b), (c) and (d) show the data/model residuals for models 1,2 and 3 respectively.}
\label{fig:psd1}
\end{figure}

\begin{figure*}
\mbox{\subfigure{\includegraphics[width=0.35\textwidth,angle=90]{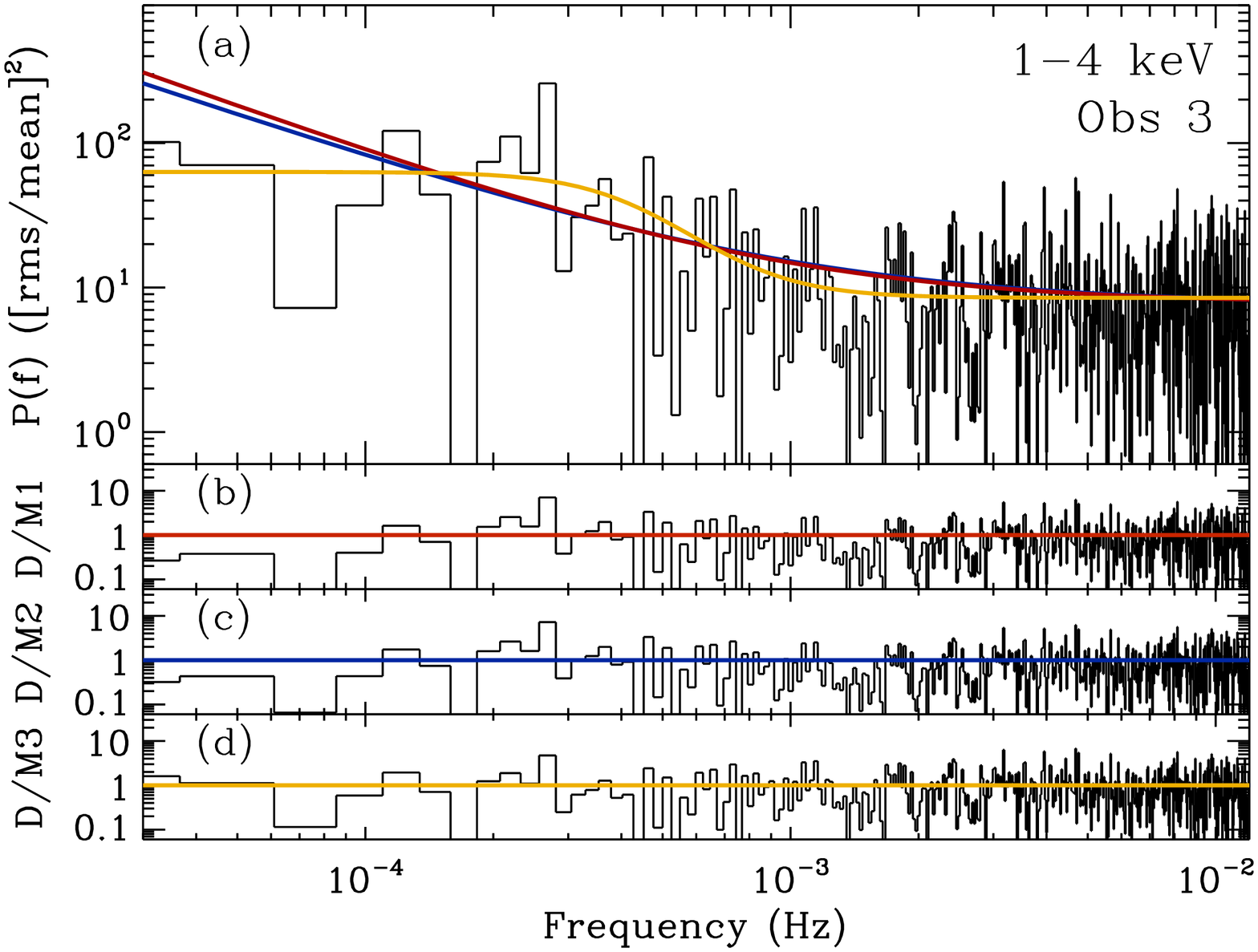}}
\hspace{15pt}
\subfigure{\includegraphics[width=0.35\textwidth,angle=90]{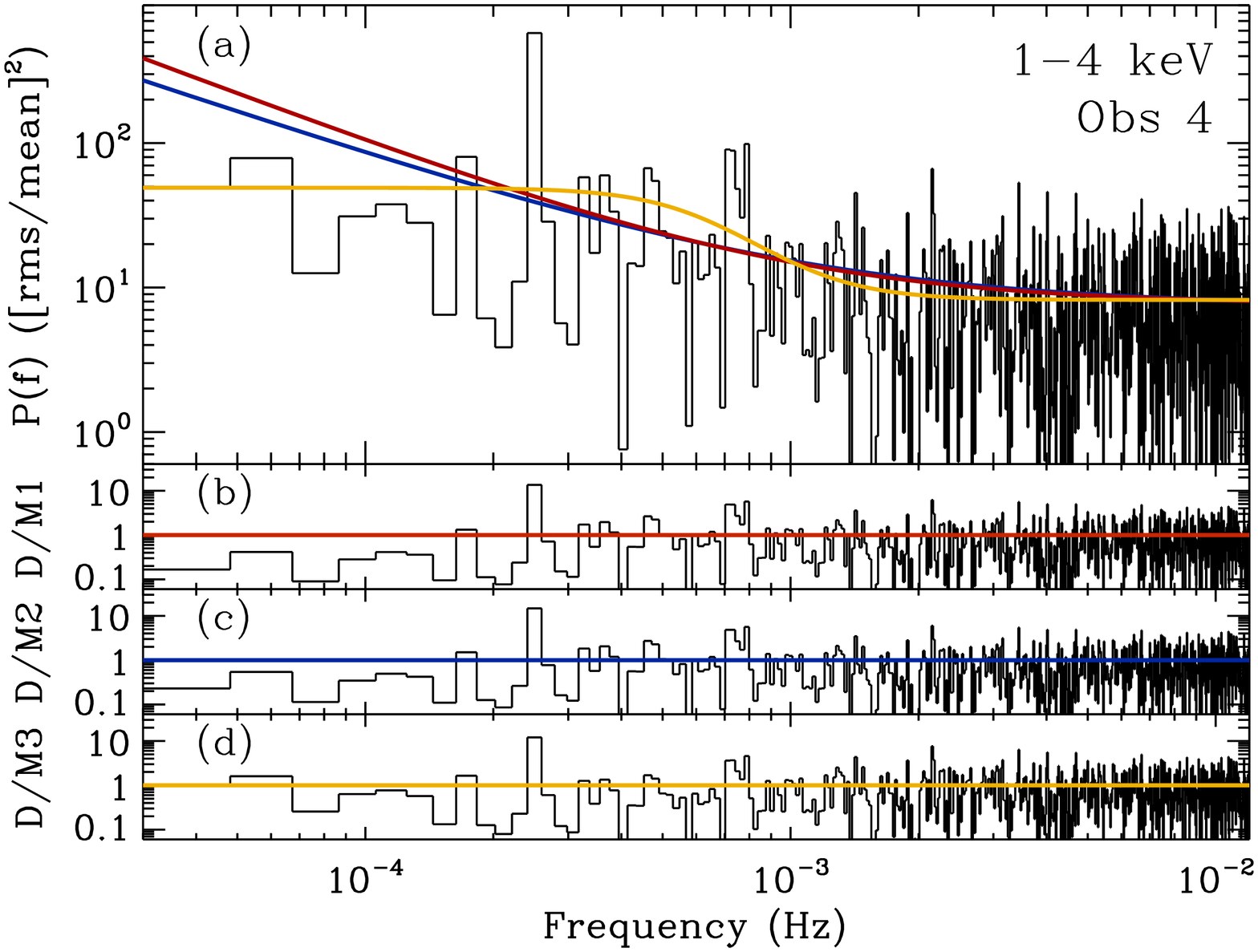}}
}\\
\mbox{\subfigure{\includegraphics[width=0.35\textwidth,angle=90]{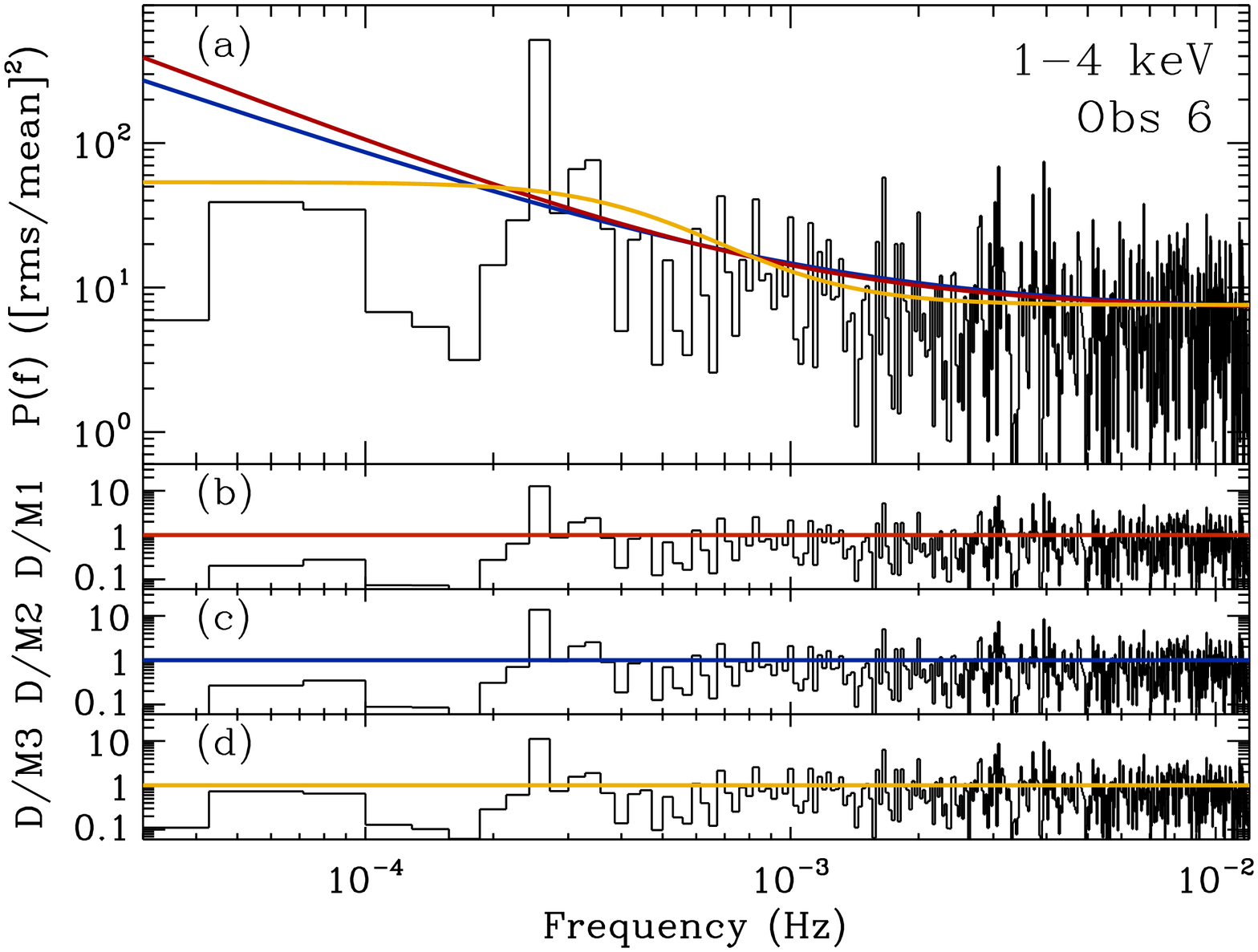}}
\hspace{15pt}
\subfigure{\includegraphics[width=0.35\textwidth,angle=90]{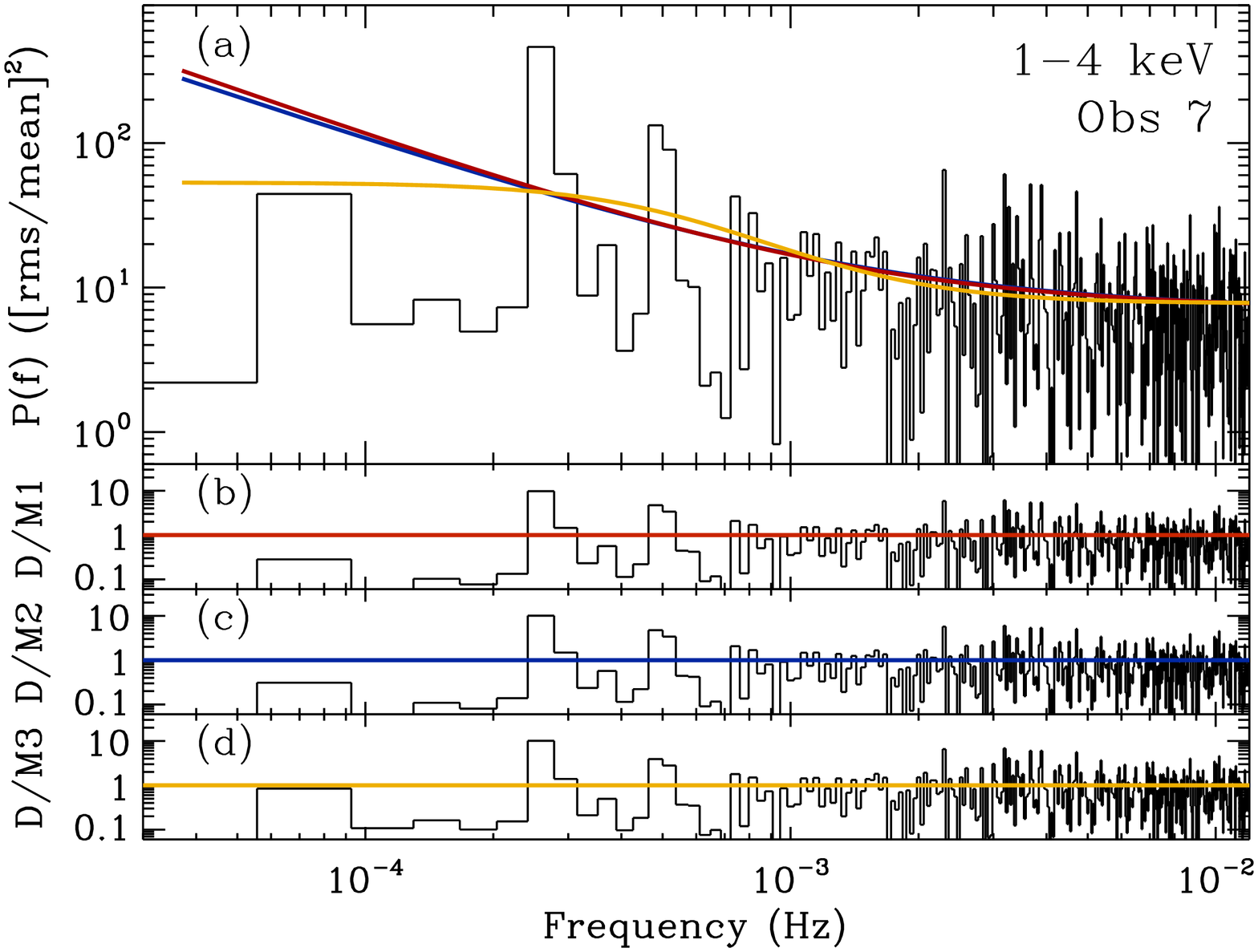}}
}\\
\caption{Obs3,4,6 and 7 data and PSD model fits for the 1.0--4.0\,keV (hard) band are shown in panel (a), for Model 1 (red), Model 2 (blue) and Model 3 (yellow).  Panels (b), (c) and (d) show the data/model residuals for models 1, 2 and 3 respectively.}
\label{fig:psd}
\end{figure*}

We estimated the PSD using the standard method of calculating the periodogram (e.g. \citealt{priestley81}; \citealt{PercivalWalden93}), using a $\rmn{[rms/mean]}^2$ PSD normalisation (e.g. \citealt{vaughan03a}).  We estimated the periodogram in three energy bands; 0.3--0.8\,keV (soft), 1.0--4.0\,keV (hard) and 0.2--10.0\,keV (total).  The hard band was chosen to maximise the contribution of the power-law spectral component (e.g. M11; \citeauthor[][{\it in prep}]{Markeviciute14prep}) whilst maintaining a high signal-to-noise ratio (S/N) light curve.  The hard-band PSDs for Obs1,3,4,6 and 7 are shown in Figs.~\ref{fig:psd1}--\ref{fig:psd}.

The PSDs were fitted using the maximum likelihood method of V10.  The aim of the fitting procedure is to distinguish between continuum models before testing the preferred continuum model for deficiencies that indicate the presence of a narrow coherent feature.  We refer the reader to V10 (and references therein) for a full discussion.

The best-fitting model parameters were found by maximizing the joint likelihood function (eq. 16 of V10).  This is equivalent to minimizing (twice) the minus log-likelihood function, which is performed using Markov Chain Monte Carlo (MCMC) simulations from the posterior distribution.  In general, 5 chains of 50,000 iterations were used to ensure convergence.

From the simulated parameter draws we form the posterior predictive distribution (V10, Appendix B), which is used to find the distribution of a test statistic and the associated posterior predictive $p$-value (ppp).  We use 5000 simulations to form the predictive distribution.  A likelihood ratio test (LRT) statistic (eq. 22 of V10) can be used for comparing nested models (e.g. \citealt{ProtassovETAL02}; V10) and was used to select between continuum models, $H_1$ and $H_0$.  The \textit{null}-hypothesis model $H_0$ was rejected when $p_{\rm LRT} < 0.01$, which is considered a conservative threshold (V10).

Once the continuum model has been selected we investigate the residuals for potential QPOs using two test statistics.  To assess the overall model fit we use the summed square error, $T_{\rm SSE}$, which is based on the traditional chi-sq statistic (eq. 21 of V10).  A small $p_{\rm SSE}$ indicates an inadequacy in the continuum modelling.  To investigate significant outliers we use $T_{\rm R} = {\rm max}_j \hat{R}_j$, where $\hat{R} = 2I_{j} / \hat{S}_j$ and $I_j$ is the observed periodogram and $S_j$ is the model power spectrum at frequency $\nu_j$.  A small $p_{\rm R}$ indicates that the largest outlier is unusual under the best-fitting continuum model.

We used two simple continuum models that have been used extensively in the literature to fit AGN PSDs (e.g. \citealt{uttley02a}; \citealt{VaughanFabian03}; \citealt{vaughan03a}; \citealt{vaughan03b}; \citealt{mchardy04}; \citealt{GonzalezMartinVaughan12}).  The simplest model (Model 1) is a power law plus constant:

\begin{equation}
\label{eqn:pl}
   P(\nu) = N \nu^{- \alpha} + C
\end{equation}

\noindent with three free parameters, where $N$ is the power-law normalisation.  The second model is a bending power-law (e.g. \citealt{mchardy04}):

\begin{equation}
\label{eqn:bendpl}
   P(\nu) = \frac{N \nu^{{\alpha}_{\rm low}}}{1 + (\nu / \nu_{\rm bend})^{{\alpha}_{\rm low}-{{\alpha}_{\rm high}}}} + C
\end{equation}

\noindent with 4 free parameters, where $\nu_{\rm bend}$ is the bend frequency, ${\alpha}_{\rm low}$ and ${\alpha}_{\rm high}$ describe the slope below and above $\nu_{\rm bend}$ respectively.  In both models, $C$ is a non-negative, additive constant which accounts for the Poisson noise level in the fitting process.  We use two versions of this model: one with ${\alpha}_{\rm low} = 1$ (Model 2), and one with ${\alpha}_{\rm low} = 0$ (Model 3).  A typical value of ${\alpha}_{\rm low} = 1$ is found from long-term X-ray monitoring studies (e.g. \citealt{uttley02a}; \citealt{MarkowitzETAL03}; \citealt{mchardy04}; \citealt{GonzalezMartinVaughan12}), however we test ${\alpha}_{\rm low} = 0$ as significances of QPO features are strongly dependent on the continuum modelling (e.g. \citealt{VaughanUttley05, VaughanUttley06}).  When fitting for Model 1 we take Model 2 as $H_0$ for the LRT.  When fitting for Models 2 and 3 we used Model 1 as $H_0$, despite now having non-nested models in the LRT.  However, the reference distribution of the LRT is constructed using MCMC draws of the posterior density (e.g. \citealt{Williams1970}).  We initially set $\nu_{\rm bend} = 2.6 \times 10^{-4}$\,Hz and all model parameters have large dispersion around the mean level in the prior distributions (see V10, section 9.4).


\subsection{Results}

For the total-band PSDs, only Obs1 showed a significant outlier in the continuum modelling, with Model 2 preferred ($p_{\rmn{LRT}} = 0.004$; $p_{\rmn{SSE}} = 0.036$; $p_{\rmn R} = 0.041$), in agreement with the findings in G08 and V10.  The lack of detection in the latter observations is consistent with M11.  As we are only interested in QPO detections here, we do not show the best fit parameters, however these are consistent with the values reported for the 0.2--10\,keV band in V10 and \citep{GonzalezMartinVaughan12}.  The soft band revealed no detection of a narrow coherent feature, with the exception of of Obs1 ($p_{\rmn{SSE}} = 0.015$; $p_{\rmn R} = 0.017$), where Model 1 is the preferred continuum model.

The model fitting results for the hard band for all observations are shown in Table~\ref{fitresults} and the PSD with model fits for each of the low flux/spectrally-harder observations (Obs1,3,4,6,7) is shown in Figs.~\ref{fig:psd1}--\ref{fig:psd}.  The LRT ppp checks clearly show that Model 1 is the preferred continuum model in Obs1,3,4,6,7 (see column 3 of Table~\ref{fitresults}).  These five observations show deficiencies in the continuum modelling, strongly indicating the presence of a narrow coherent feature.  The best fitting Model 1 to Obs1,3,4,6,7 has a consistent shape, with mean $\alpha \sim 1.16$.  Despite continuum Model 1 being preferred in these five observations, we also present the fit results for Models 2 and 3 in Table~\ref{fitresults} and show the best fitting model and residuals in Figs.~\ref{fig:psd1}--\ref{fig:psd}.  For these continuum models, the narrow coherent features remain significant.

Obs0 shows a very weak outlier ($2I/S = 9$) at the QPO frequency, however $p_{\rmn{SSE}} \sim p_{\rmn R} \sim 0.15$, for the three continuum models.  This very weak detection is most likely due to the short observation length.  The hard-band PSDs of Obs2 and 5 show no narrow coherent features.  The normalisation of the broadband noise component in Obs2 and 5 is significantly weaker than that observed in Obs1,3,4,6,7 and the Poisson noise dominates at the QPO frequency.

The most significant model outliers ($2I/S$, see column 6 of Table~\ref{fitresults}) occur at the same frequency (within the frequency resolution of the data) in all five of these observations.  As was observed in G08, the QPO lies within one periodogram frequency bin, making the QPO highly coherent.  The quality factor $Q = \nu / \Delta \nu$ is 24, 11, 13, 9 and 7 for Obs1,3,4,6,7 respectively.  A $Q \gsim 5$ is typically considered as a coherent feature.

The rms fractional variability in the QPO is given in column 7 of Table~\ref{fitresults}, with a mean of $\sim 10$ per cent, consistent with G08 and \citealt{MiddletonETAL09} above 1\,keV.  The rms of Obs1 is lower than the latter observations, which could be due to the splitting of the QPO power across neighbouring frequencies.  Indeed, Obs1 (Fig.~\ref{fig:psd1}) shows another strong peak in the PSD at $2.4 \times 10^{-4}$, just below the QPO frequency.  This corresponds to a period $P = 4090$\,s, which was also observed in \citet{HuETAL14} using an alternative method involving the Hilbert-Huang transform.  If indeed the QPO is split across three frequencies in Obs1 then we now have $Q \sim 8$, consistent with the remaining observations.

HFQPOs in BHBs often display 3:2 or 2:1 harmonic structures in their power spectra (e.g. \citealt{remillard2002, remillard2003prec}; \citealt{RemillardMcClintock06}).  Obs7 (Fig.~\ref{fig:psd}) shows outliers in two frequency bins at $\sim 5 \times 10^{-4}$\,Hz, $\sim$ 2 times the QPO frequency, with $Q \sim 7$.  With $2I/S \sim 9$ we can rule out this feature being a statistical fluctuation with $\sim 98$ per cent confidence (eq. 16 of \citealt{vaughan05a}), or $\sim 2.5 \sigma$.  This feature is explored in \citeauthor[][{\it in prep}]{Markeviciute14prep}.

\begin{table}
\caption{Results of model fits to the hard band (1.0--4.0\,keV) PSD .  Column (1) shows the observation number, column (2) the frequency of the most significant model outlier (interpreted as the QPO), column (3) shows the posterior predictive $p$-values (ppp) for the LRT between the \textit{null} hypothesis and alternative hypothesis.  Columns (4) and (5) show the ppp for $T_{\rmn{SSE}}$ and  $T_{\rmn R}$ respectively.  Column (6) shows the highest data/model outlier, corresponding to the frequency given in column (2).  See Sec.~\ref{sec:psd} for model and test statistic details.}
\begin{tabular}{lcccccc}
\hline
Obs  & $f_{\rmn{QPO}}$  & $p_{\rmn{LRT}}$ & $p_{\rmn{SSE}}$  & $p_{\rmn R}$ &   $2I_j/S_j$ & rms(QPO) \\
\vspace{-0.3cm}\\
    & $\times 10^{-4}$&  \multicolumn{4}{c}{} & $\%$ \\
(1) & (2) &(3) &(4) &(5) &(6) & (7) \\ 
\hline
    & &  \multicolumn{5}{c}{Model 1}  \\
1   & 2.7  & 0.0002  & 0.004  & 0.0004  & 32.2 & 7  \\
2   &   -    & 0.8078  & 0.686  & 0.6304  &   -  & -  \\
3   & 2.7  & 0.1768  & 0.034  & 0.0539  & 14.3 & 8  \\
4   & 2.5  & 0.0011  & 0.001  & 0.0014  & 27.1 & 10 \\
5   &    -   & 0.9898  & 0.215  & 0.8168  &    - &  - \\
6   & 2.6  & 0.0024  & 0.003  & 0.0022  & 25.3 & 12 \\
7   & 2.6  & 0.0021  & 0.022  & 0.0354  & 19.5 & 13 \\
\hline
    & &  \multicolumn{5}{c}{Model 2}  \\
1   & 2.7  & 0.94  & 0.0006 & 0.0006 & 35.7 &  7  \\
2   &   -    & 0.59  & 0.6661 & 0.6304 & -    &  -  \\
3   & 2.7  & 0.37  & 0.0296 & 0.4988 & 14.4 &  8  \\
4   & 2.5  & 0.89  & 0.0004 & 0.0004 & 29.5 &  10 \\
5   &    -   & 0.43  & 0.2148 & 0.8168 &  -   &  -  \\
6   & 2.6  & 0.73  & 0.0014 & 0.0004 & 27.5 &  12 \\
7   & 2.6  & 0.72  & 0.0194 & 0.0216 & 20.2 &  13 \\
\hline
    & &  \multicolumn{5}{c}{Model 3}  \\
1   & 2.7  & 0.32 & 0.023 & 0.0111 & 19.6 & 7  \\
2   &   -    & 0.98 & 0.614 & 0.5952 & -    & -  \\
3   & 2.7  & 0.74 & 0.067 & 0.6728 & 13.6 & 8  \\
4   & 2.5  & 0.53 & 0.003 & 0.0056 & 24.2 & 10 \\
5   &    -   & 0.34 & 0.238 & 0.8568 & -    & -  \\
6   & 2.6  & 0.50 & 0.002 & 0.0102 & 22.4 & 12 \\
7   & 2.6  & 0.59 & 0.009 & 0.0174 & 20.3 & 13 \\
\hline
\end{tabular}
\label{fitresults}
\end{table}

\section{Discussion and Conclusions}
\label{sec:disco}

We show that a significant narrow coherent feature at $\sim 2.6 \times 10^{-4}$\,Hz in the PSD, most likely associated with a QPO, is present in five observations of \rej.  This increases the duration where a QPO is confirmed as significantly present in \rej~by a factor three, totalling $\sim 250$\,ks.  The QPO is only detected when the source has a low flux and is spectrally harder (6 out of the 8 \xmmn observations).  The QPO is detected at all energies in Obs1, however it is only detected in the hard band in the remaining low-flux observations, showing a clear energy dependence of this feature.  The QPO shows remarkable stability with a frequency that has remained constant over four years (and possibly 9 years).  The significant detection of the QPO at the same frequency in five independent observations confirms unambiguously the presence of the QPO feature in this source. 

The QPO was originally detected in the 0.3--10.0\,keV band in G08 (our Obs1).  This was subsequently confirmed by V10 using an identical analysis presented here.  The 0.3--10.0\,keV PSD of some further \xmmn observations of \rej~(our Obs2,3,4) were analysed in M11.  No evidence for the QPO feature at any frequency was found, consistent with our total band PSD results for these observations.  

The black hole mass in \rej~is highly uncertain, though best estimates are converging on $\sim 1-4 \times 10^{6} \Msun$ (e.g. \citealt{BianHuang10}; \citealt{JinETAL11}).  M10 and M11 proposed that the QPO in \rej~was analogous with the 67 mHz QPO in GRS 1915+105, based on the inferred $L_{\rm Bol}/\Ledd \gsim 1$, and mass scaling.  Recently, however, \citet{Mendez13} observed a soft lag, where softer photons are delayed with respect to harder photons, in the 35 mHz HFQPO in GRS 1915+105 (\citealt{BelloniAltamirano13}).  A soft lag is also observed at the QPO frequency in \rej~(\citealt{zoghbi11b}), leading \citet{Mendez13} to suggest the QPO in \rej~is an analogue of the 35 mHz QPO in GRS 1915+105.  We would also then expect to see the analogue of the 67 mHz QPO in \rej, which we could be seeing in Obs7 (Fig.~\ref{fig:psd}).

The QPO in \rej~is highly coherent ($Q \sim 10$), consistent with values of HFQPOs observed in BHBs (e.g. \citealt{Mendez13}). \citet{Mendez13} find the 35 and 67 mHz QPOs in GRS 1915+105 have $Q \sim 9$ and $\sim 33$, respectively.  The large $Q$ observed in the 67 mHz harmonic contrasts with the $Q \sim 7$ observed in Obs7.  This may be due to this being an unrelated feature, or that this harmonic feature is dependent on some other system property, such as spectral hardness.

A clear energy dependence to the QPO component is seen in the low-flux/spectrally-harder observations.  The energy dependence of the QPO suggests the feature is generated in the power-law spectral component (which dominates above $\sim 1$\,keV), as was suggested in M11.  Any hardening of this component will then affect the detectability of the QPO at soft energies in these observations.  

The disappearance of the QPO in the high-flux/spectrally-softer observations (Obs2 and 5) could be related to the increase in the soft spectral component (M11).  This component dominates over the power law up to $\sim 2$\,keV and a lower fractional rms is observed in these observations as a result (M11); this then makes detecting the QPO difficult given the data quality, as we observe in the hard-band PSD of Obs2 and 5.  Alternatively, the disappearance of the QPO is physically related to the increase in the soft component.

AGN offer a better opportunity to understand the HFQPO phenomenon, due to having higher counts per characteristic timescale.  The firm detection of the QPO in multiple observations of \rej~allows us to study this feature in more detail.  In a follow up paper (\citeauthor[][{\it in prep}]{Markeviciute14prep}) we explore several other aspects of the spectral timing in \rej~and discuss these results in the context of current models for the origin of HFQPOs.


\section*{Acknowledgements}

We thank the anonymous referee for a thorough reading of the manuscript and providing insightful comments.  We acknowledge useful discussions with Simon Vaughan, Phil Uttley, Lucy Heil and Adam Ingram.  WNA, EK and ACF acknowledge support from the European Union Seventh Framework Programme (FP7/2013--2017) under grant agreement n.312789, StrongGravity.  This paper is based on observations obtained with \xmm, an ESA science mission with instruments and contributions directly funded by ESA Member States and the USA (NASA). 

\footnotesize{
\bibliographystyle{mn2e}
\bibliography{rej1034qpo}

\begin{thebibliography}{}

\bibitem[\protect\citeauthoryear{{Altamirano} \& {Belloni}}{{Altamirano} \&
  {Belloni}}{2012}]{Altamirano2012}
{Altamirano} D.,  {Belloni} T.,  2012, \apjl, 747, L4

\bibitem[\protect\citeauthoryear{{Ballet}}{{Ballet}}{1999}]{ballet99}
{Ballet} J.,  1999, \aaps, 135, 371

\bibitem[\protect\citeauthoryear{{Belloni} \& {Altamirano}}{{Belloni} \&
  {Altamirano}}{2013}]{BelloniAltamirano13}
{Belloni} T.~M.,  {Altamirano} D.,  2013, \mnras, 432, 19

\bibitem[\protect\citeauthoryear{{Bian} \& {Huang}}{{Bian} \&
  {Huang}}{2010}]{BianHuang10}
{Bian} W.-H.,  {Huang} K.,  2010, \mnras, 401, 507

\bibitem[\protect\citeauthoryear{{Czerny}, {Lachowicz}, {Dov{\v c}iak},
  {Karas}, {Pech{\'a}{\v c}ek} \& {Das}}{{Czerny}
  et~al.}{2010}]{CzernyETAL2010}
{Czerny} B.,  {Lachowicz} P.,  {Dov{\v c}iak} M.,  {Karas} V.,  {Pech{\'a}{\v
  c}ek} T.,    {Das} T.~K.,  2010, \aap, 524, A26

\bibitem[\protect\citeauthoryear{{Czerny}, {Lachowicz}, {Dov{\v c}iak},
  {Karas}, {Pech{\'a}{\v c}ek} \& {Das}}{{Czerny} et~al.}{2012}]{Czerny12}
{Czerny} B.,  {Lachowicz} P.,  {Dov{\v c}iak} M.,  {Karas} V.,  {Pech{\'a}{\v
  c}ek} T.,    {Das} T.~K.,  2012, Journal of Physics Conference Series, 372,
  012055

\bibitem[\protect\citeauthoryear{{Davis}}{{Davis}}{2001}]{davis01}
{Davis} J.~E.,  2001, \apj, 562, 575

\bibitem[\protect\citeauthoryear{{Gierli{\'n}ski}, {Middleton}, {Ward} \&
  {Done}}{{Gierli{\'n}ski} et~al.}{2008}]{gierlinski08}
{Gierli{\'n}ski} M.,  {Middleton} M.,  {Ward} M.,    {Done} C.,  2008, \nat,
  455, 369

\bibitem[\protect\citeauthoryear{{Gonz{\'a}lez-Mart{\'{\i}}n} \&
  {Vaughan}}{{Gonz{\'a}lez-Mart{\'{\i}}n} \&
  {Vaughan}}{2012}]{GonzalezMartinVaughan12}
{Gonz{\'a}lez-Mart{\'{\i}}n} O.,  {Vaughan} S.,  2012, \aap, 544, A80

\bibitem[\protect\citeauthoryear{{Hu}, {Chou}, {Yang} \& {Su}}{{Hu}
  et~al.}{2014}]{HuETAL14}
{Hu} C.-P.,  {Chou} Y.,  {Yang} T.-C.,    {Su} Y.-H.,  2014, \apj, 788, 31

\bibitem[\protect\citeauthoryear{{Jin}, {Ward}, {Done} \& {Gelbord}}{{Jin}
  et~al.}{2011}]{JinETAL11}
{Jin} C.,  {Ward} M.~J.,  {Done} C.,    {Gelbord} J.,  2011, in American
  Astronomical Society Meeting Abstracts \#218 {An Optical and X-ray Spectral
  Study of Unobscured AGN: The SED Model Fits}.
p.~408

\bibitem[\protect\citeauthoryear{{Lin}, {Irwin}, {Godet}, {Webb} \&
  {Barret}}{{Lin} et~al.}{2013}]{LinETAL13}
{Lin} D.,  {Irwin} J.~A.,  {Godet} O.,  {Webb} N.~A.,    {Barret} D.,  2013,
  \apjl, 776, L10

\bibitem[\protect\citeauthoryear{{Markevi\v{c}i\={u}t\.{e}}, {Alston}, {Kara},
  {Fabian} \& {Middleton}}{{Markevi\v{c}i\={u}t\.{e}}
  et~al.}{2014}]{Markeviciute14prep}
{Markevi\v{c}i\={u}t\.{e}} J.,  {Alston} W.,  {Kara} E.,  {Fabian} A.~C.,
  {Middleton} M.,  2014, \mnras,~{\it in prep}

\bibitem[\protect\citeauthoryear{{Markowitz}, {Edelson}, {Vaughan}, {Uttley},
  {George}, {Griffiths}, {Kaspi}, {Lawrence}, {McHardy}, {Nandra}, {Pounds},
  {Reeves}, {Schurch} \& {Warwick}}{{Markowitz} et~al.}{2003}]{MarkowitzETAL03}
{Markowitz} A.,  {Edelson} R.,  {Vaughan} S.,  {Uttley} P.,  {George} I.~M.,
  {Griffiths} R.~E.,  {Kaspi} S.,  {Lawrence} A.,  {McHardy} I.,  {Nandra} K.,
  {Pounds} K.,  {Reeves} J.,  {Schurch} N.,    {Warwick} R.,  2003, \apj, 593,
  96

\bibitem[\protect\citeauthoryear{{McHardy}, {Papadakis}, {Uttley}, {Page} \&
  {Mason}}{{McHardy} et~al.}{2004}]{mchardy04}
{McHardy} I.~M.,  {Papadakis} I.~E.,  {Uttley} P.,  {Page} M.~J.,    {Mason}
  K.~O.,  2004, \mnras, 348, 783

\bibitem[\protect\citeauthoryear{{M{\'e}ndez}, {Altamirano}, {Belloni} \&
  {Sanna}}{{M{\'e}ndez} et~al.}{2013}]{Mendez13}
{M{\'e}ndez} M.,  {Altamirano} D.,  {Belloni} T.,    {Sanna} A.,  2013, \mnras,
  435, 2132

\bibitem[\protect\citeauthoryear{{Middleton} \& {Done}}{{Middleton} \&
  {Done}}{2010}]{MiddletonDone10}
{Middleton} M.,  {Done} C.,  2010, \mnras, 403, 9

\bibitem[\protect\citeauthoryear{{Middleton}, {Done}, {Ward}, {Gierli{\'n}ski}
  \& {Schurch}}{{Middleton} et~al.}{2009}]{MiddletonETAL09}
{Middleton} M.,  {Done} C.,  {Ward} M.,  {Gierli{\'n}ski} M.,    {Schurch} N.,
  2009, \mnras, 394, 250

\bibitem[\protect\citeauthoryear{{Middleton}, {Uttley} \& {Done}}{{Middleton}
  et~al.}{2011}]{middletonetal11}
{Middleton} M.,  {Uttley} P.,    {Done} C.,  2011, \mnras, 417, 250

\bibitem[\protect\citeauthoryear{{Mushotzky}, {Done} \& {Pounds}}{{Mushotzky}
  et~al.}{1993}]{mushotzky93}
{Mushotzky} R.~F.,  {Done} C.,    {Pounds} K.~A.,  1993, \araa, 31, 717

\bibitem[\protect\citeauthoryear{Percival \& Walden}{Percival \&
  Walden}{1993}]{PercivalWalden93}
Percival D.,  Walden A.,  1993, Spectral Analysis for Physical Applications.
Cambridge University Press

\bibitem[\protect\citeauthoryear{Priestley}{Priestley}{1981}]{priestley81}
Priestley M.,  1981, Spectral analysis and time series.
Academic Press, London

\bibitem[\protect\citeauthoryear{{Protassov}, {van Dyk}, {Connors}, {Kashyap}
  \& {Siemiginowska}}{{Protassov} et~al.}{2002}]{ProtassovETAL02}
{Protassov} R.,  {van Dyk} D.~A.,  {Connors} A.,  {Kashyap} V.~L.,
  {Siemiginowska} A.,  2002, \apj, 571, 545

\bibitem[\protect\citeauthoryear{{Reis}, {Miller}, {Reynolds}, {G{\"u}ltekin},
  {Maitra}, {King} \& {Strohmayer}}{{Reis} et~al.}{2012}]{ReisETAL13}
{Reis} R.~C.,  {Miller} J.~M.,  {Reynolds} M.~T.,  {G{\"u}ltekin} K.,  {Maitra}
  D.,  {King} A.~L.,    {Strohmayer} T.~E.,  2012, Science, 337, 949

\bibitem[\protect\citeauthoryear{{Remillard}, {Muno}, {McClintock} \&
  {Orosz}}{{Remillard} et~al.}{2003}]{remillard2003prec}
{Remillard} R.,  {Muno} M.,  {McClintock} J.~E.,    {Orosz} J.,  2003, in
  {Durouchoux} P.,  {Fuchs} Y.,   {Rodriguez} J.,  eds, New Views on
  Microquasars {X-ray QPOs in black-hole binary systems}.
p.~57

\bibitem[\protect\citeauthoryear{{Remillard} \& {McClintock}}{{Remillard} \&
  {McClintock}}{2006}]{RemillardMcClintock06}
{Remillard} R.~A.,  {McClintock} J.~E.,  2006, \araa, 44, 49

\bibitem[\protect\citeauthoryear{{Remillard}, {Morgan}, {McClintock}, {Bailyn}
  \& {Orosz}}{{Remillard} et~al.}{1999}]{Remillard1999}
{Remillard} R.~A.,  {Morgan} E.~H.,  {McClintock} J.~E.,  {Bailyn} C.~D.,
  {Orosz} J.~A.,  1999, \apj, 522, 397

\bibitem[\protect\citeauthoryear{{Remillard}, {Muno}, {McClintock} \&
  {Orosz}}{{Remillard} et~al.}{2002}]{remillard2002}
{Remillard} R.~A.,  {Muno} M.~P.,  {McClintock} J.~E.,    {Orosz} J.~A.,  2002,
  \apj, 580, 1030

\bibitem[\protect\citeauthoryear{{Shakura} \& {Sunyaev}}{{Shakura} \&
  {Sunyaev}}{1973}]{shaksuny73}
{Shakura} N.~I.,  {Sunyaev} R.~A.,  1973, \aap, 24, 337

\bibitem[\protect\citeauthoryear{{Strohmayer}}{{Strohmayer}}{2001}]{Strohmayer2001}
{Strohmayer} T.~E.,  2001, \apjl, 552, L49

\bibitem[\protect\citeauthoryear{{Str{\"u}der}, {Briel}, {Dennerl} \&
  {Hartmann}}{{Str{\"u}der} et~al.}{2001}]{struder01}
{Str{\"u}der} L.,  {Briel} U.,  {Dennerl} K.,    {Hartmann} R.,  2001, \aap,
  365, L18

\bibitem[\protect\citeauthoryear{{Uttley}, {McHardy} \& {Papadakis}}{{Uttley}
  et~al.}{2002}]{uttley02a}
{Uttley} P.,  {McHardy} I.~M.,    {Papadakis} I.~E.,  2002, \mnras, 332, 231

\bibitem[\protect\citeauthoryear{{van der Klis}}{{van der
  Klis}}{2006}]{vanderklis2006}
{van der Klis} M.,  2006, in {Lewin}, W.~H.~G. and {van der Klis}, M., eds.,
  Compact stellar X-ray sources.
Cambridge University Press, Cambridge, UK, p.~39

\bibitem[\protect\citeauthoryear{{Vaughan}}{{Vaughan}}{2005}]{vaughan05a}
{Vaughan} S.,  2005, \aap, 431, 391

\bibitem[\protect\citeauthoryear{{Vaughan}}{{Vaughan}}{2010}]{vaughan10}
{Vaughan} S.,  2010, \mnras, 402, 307

\bibitem[\protect\citeauthoryear{{Vaughan}, {Edelson}, {Warwick} \&
  {Uttley}}{{Vaughan} et~al.}{2003}]{vaughan03a}
{Vaughan} S.,  {Edelson} R.,  {Warwick} R.~S.,    {Uttley} P.,  2003, \mnras,
  345, 1271

\bibitem[\protect\citeauthoryear{{Vaughan} \& {Fabian}}{{Vaughan} \&
  {Fabian}}{2003}]{VaughanFabian03}
{Vaughan} S.,  {Fabian} A.~C.,  2003, \mnras, 341, 496

\bibitem[\protect\citeauthoryear{{Vaughan}, {Fabian} \& {Nandra}}{{Vaughan}
  et~al.}{2003}]{vaughan03b}
{Vaughan} S.,  {Fabian} A.~C.,    {Nandra} K.,  2003, \mnras, 339, 1237

\bibitem[\protect\citeauthoryear{{Vaughan} \& {Uttley}}{{Vaughan} \&
  {Uttley}}{2005}]{VaughanUttley05}
{Vaughan} S.,  {Uttley} P.,  2005, \mnras, 362, 235

\bibitem[\protect\citeauthoryear{{Vaughan} \& {Uttley}}{{Vaughan} \&
  {Uttley}}{2006}]{VaughanUttley06}
{Vaughan} S.,  {Uttley} P.,  2006, Advances in Space Research, 38, 1405

\bibitem[\protect\citeauthoryear{{Wilkinson} \& {Uttley}}{{Wilkinson} \&
  {Uttley}}{2009}]{WilkinsonUttley09}
{Wilkinson} T.,  {Uttley} P.,  2009, \mnras, 397, 666

\bibitem[\protect\citeauthoryear{Williams}{Williams}{1970}]{Williams1970}
Williams D.~A.,  1970, Biometrics, 26, pp. 23

\bibitem[\protect\citeauthoryear{{Zoghbi} \& {Fabian}}{{Zoghbi} \&
  {Fabian}}{2011}]{zoghbi11b}
{Zoghbi} A.,  {Fabian} A.~C.,  2011, \mnras, 418, 2642

\end{thebibliography}
}


\label{lastpage}


\end{document}